\documentstyle[12pt,psfig,draft]{nature-ejb}

\newcommand{\nature}{\mbox{\it Nature }}
\newcommand{\et}{\mbox{\it et al. }}
\newcommand{\ApJ}{\mbox{\it Astrophys. J. }}

\newcommand{\AAP}{\mbox{\it Astron. Astrophys. }}

\input{psfig.sty}
\begin{document}

%\psdraft

\title{\Large \bf Thermal Emission from a Hot Cocoon Surrounding the
Jet of XRF 060218}

\author{
Enwei Liang \affiliation[1]{Department of Physics,
 University of Nevada, Las Vegas, NV 89154,
USA}$^,$ \affiliation[2]{Department of Physics,
 Guangxi University, Nanning 530004, China},
Bing Zhang \affiliationmark[1],
Bin-Bin Zhang \affiliation[3]{National Astronomical
Observatories/Yunnan Observatory, Chinese Academy of Sciences, Kunming
650011, China}$^,$
\affiliation[4]{The Graduate School of the Chinese Academy of
Sciences, Beijing 100039, China},
\& Z. G. Dai \affiliation[5]{Department of Astronomy, Nanjing University,
Nanjing 210093, China }
}

\dates{\today}{}
\headertitle{} \mainauthor{}
\smallskip

\summary{It is long speculated that long duration gamma-ray bursts (GRBs) originate from
a relativistic jet emerging from a collapsing massive star
progenitor\cite{Woosley93,Paczynski98,MacFadyen99,Zhang03}. Although associations of
core-collapsing supernovae with long GRB afterglows have been identified in a number of
systems\cite{Galama98,Kulkarni98,Stanek03,Hjorth03}, including the latest X-ray flash
(XRF) 060218/SN 2006aj connection detected by {\em
Swift}\cite{Campana06,Modjaz06,Pian06,Sollerman06,Mirabal06,Cobb06,Soderberg06}, direct
evidence of a relativistic jet emerging from a collapsing star is still lacking. Here we
report the detection of a thermal emission component (high-$T$ component) accompanying
the prompt X-ray emission of XRF 060218, with temperature $kT_H=1.21^{+0.22}_{-0.24}$ keV
and effective blackbody radius $R_H\sim 5\times 10^{9}~{\rm cm}$. This high-$T$ component
co-exists with another low-$T$ thermal component as reported by Campana et al.
\cite{Campana06} for at least 2700 seconds, but evolves independently with respect to the
low-$T$ component by tracing the lightcurve of the non-thermal component. We identify
this high-$T$ thermal component as the emission of a hot cocoon surrounding the
relativistic jet, as expected from the theoretical
models\cite{Woosley99,Meszaros01,Ramirez02,Zhang03,Waxman03}.}

\maketitle \noindent XRF 060218 was detected with the Swift/Burst Alert Telescope (BAT)
on 2006 February 18.149 UT. Swift slewed autonomously to the burst and the X-ray
telescope (XRT) began to collect data 159 s after the burst trigger. We analyze the XRT
data in the first orbit ($t=159-2740$ seconds since the burst trigger). We reduce the
data with the Swift/XRT tools and process the spectrum with the Xspec software package.
We first fit the time-integrated spectrum with a combined model of a single-temperature
($kT$) black body model (BB model) normalized to its apparent area at a given distance
and a single power law ($dN(E)/dE\propto E^{-\Gamma}$; PL model), incorporating the
neutral hydrogen absorbtion in both our Galaxy and the GRB host galaxy. The observed
effective" black body radius" ($R$) could be inferred from the normalization term ($K$)
of this model, $R_{\rm BB}=K^{1/2}\times D_{10 {\rm kpc}}$ km, where $D_{10 {\rm kpc}}$
is the source distance in units of 10 kpc. The $N_H$ of our Galaxy\cite{Dickey90} for
this burst is $\sim 1.1\times 10^{21}$ cm$^{-2}$. We find that the fitting results depend
on the initial guess value of $kT$. If $kT$ is assigned an initial guess value less than
1 keV, one can get the best fit results with $kT\sim 0.12$ keV, the thermal component
identified by Campana et al.\cite{Campana06}. However, we also find marginally acceptable
fitting results with a high-$T$ component with $kT\sim 1.3$ keV, if the initial guess
value of $kT$ is assigned greater than $1$ keV. We replace the single power law component
with a broken power law model (BKNPL model) in our fitting, and find that the high-$T$
component stands out and the two photon indices of the broken power law are $\Gamma_1
\sim -1$ (fixed) and $\Gamma_2 \sim 2.08$, respectively, with a break energy at $E_b\sim
0.80$ keV. The photon spectrum with $E<E_b$ roughly agrees with a black body spectrum,
being consistent with the low-$T$ component. This likely indicates that there exists two
thermal components in the XRT data, with $kT\sim 0.1$ keV and $kT\sim 1.3$ keV,
respectively. We then use a combined model with two thermal and one power law components
(BB+BB+PL model) to fit the spectrum. The comparison of the fitting results with the
BB+PL model and the BB+BKNPL model is shown in Table 1. It is found that besides the
previous reported thermal component with $kT_1=0.116^{+0.002}_{-0.003}$ (low-$T$
component)\cite{Campana06}, a new thermal component (high-$T$ component) with
$kT_2=1.21^{+0.22}_{-0.24}$ keV is indeed present in the XRT data. Figure 1 shows the
accumulated overall X-ray spectrum during the first orbit (159-2740 seconds since the
burst trigger), with the contributions from the two thermal components marked. We derive
the effective blackbody radii of the two thermal components as $R_{L}=3.2\times 10^{12}$
cm and $R_{H}=5\times 10^9$ cm, respectively, suggesting that the two thermal components
are from two distinct emission regions.

The existence of two thermal components is also supported by a time-dependent spectral
analysis of the X-ray data. The PL component derived from the BB+PL model shows an
unreasonable hard-to-soft evolution feature, with the photon index continuously
increasing from $\sim 2$ to $3.0$ since $t>800$ seconds (Fig.2). This implies a very
peculiar electron distribution. The PL photon index derived from the BB+BB+PL model, on
the other hand, does not show a significant evolution, with an average index of $2.2\pm
0.4$ (Fig.2). The unexpected softening derived in the BB+PL model is therefore an
artifact due to the contribution of the soft photons in the high-$T$ thermal component.

Our time-resolved spectral analysis reveals the evolution of the two thermal components.
Figure 3 displays the temporal evolution of the temperatures and the effective blackbody
radii of the two components. The high-$T$ component expands rapidly with $R_H \propto t$
and cools rapidly as $kT_H \propto t^{-0.87}$, while the low-$T$ component expands more
slowly with $R_L \propto t^{0.38}$ and only cools mildly as $kT_L \sim t^{-0.05}$. We
note, however, that the peak flux of the low-$T$ component is right at the low end of the
XRT energy band where the absorption is very strong. It would be difficult to robustly
measure the temperature evolution of this component. In Figure 4 we show both the
unabsorbed and absorbed lightcurves of the two thermal components along with the
non-thermal hard X-ray lightcurve in the 5-10 keV band. It is evident that the two
components evolve independently. While the observed flux of the low-$T$ component
increases with time, the flux of the high-$T$ component traces the non-thermal component.
This suggests that the high-$T$ component is directly related with the GRB jet.

Numerical simulations of a relativistic jet propagating in the stellar envelope of a
collapsing star reveals a hot cocoon surrounding the jet\cite{Zhang03}, which is the
waste heat from the jet enclosed by the dense envelope. When the jet, which may be mildly
relativistic inside the star, first penetrates through the stellar envelope, the
associated cocoon also breaks out and their thermal photons leak out. This hot thermal
ring surrounding the jet then subsequently expands, both forwardly and in sideways, as
the rest of jet bores its way out, The cocoon luminosity should roughly trace the
non-thermal emission luminosity of the jet. This picture is in good agreement with the
data of the high-$T$ thermal component. Combined with the low-$T$ component that has been
interpreted as the shock breakout emission of the global supernova \cite{Campana06}, our
result clearly suggests that there exist three components during the explosion of a dying
massive star. Besides the more isotropically expanding supernova ejecta, there are indeed
a collimated jet breaking through the envelope and producing the non-thermal emission and a
hot thermal cocoon component associated with it.

Assuming that the radius of the low-$T$ component $R_L \sim 3.2 \times 10^{12}$ cm is
also roughly the radius of the exploding star, the high-$T$ component then opens a very
small solid angle of $\Delta \Omega \sim (R_H / R_L)^2 \sim 2.4 \times 10^{-6}$. The radius
of the star may be smaller than $R_L$, but in any case, it is unlikely that this solid
angle reflects the solid angle of the relativistic jet, which should be very broad
according to the late afterglow data\cite{Soderberg06,Fan06} and
population studies\cite{Liang06,Cobb06}. A possible picture is
that the cocoon surrounding the jet forms a thin ring with a small
solid angle. This is much
narrower than the predicted values in numerical simulations\cite{Zhang03}. However, XRF
060218 has a much lower luminosity than the modelled typical collapsars. With much less
waste heat surrounding an under-luminous jet, the cocoon could be in
principle much thinner.

\begin{table*}
\caption{The comparison of the spectral fitting results with different models} \tiny
\begin{tabular}{llllllllll}
\hline \hline
Model&$\chi^2$&$N_H^{\rm host}(10^{22}$cm$^{-2}$)&$kT_1$(keV)&$K_1(10^6)$&$kT_2$(keV)&$K_2$&$\Gamma_1$&$\Gamma_2$&$E_b$\\
\hline

BB+PL&1.55&$0.67^{+0.02}_{-0.02}$ & $0.116^{+0.002}_{-0.003}$ & $6.71^{+1.86}_{-1.58}$ &
 &  & $1.97^{+0.02}_{-0.02}$&-&-\\

BB+BB+PL &1.53&$0.64^{+0.03}_{-0.03}$ & $0.119^{+0.005}_{-0.004}$ &
$4.88^{+2.22}_{-1.54}$ & $1.21^{+0.22}_{-0.24}$ & $10.7^{+9.6}_{-4.5}$ &
$2.02^{+0.08}_{-0.06}$&-&-\\
BB+BKNPL&1.52&$0.19^{+0.02}_{-0.02}$&-&-&$1.32^{+0.04}_{-0.05}$&$25.6^{+3.5}_{-3.4}$&$2.08^{+0.07}_{-0.06}$&$-1$(fixed)&$0.79^{+0.01}_{-0.01}$\\
\hline
\end{tabular}
\end{table*}

\newpage

{\bf Figure 1} The observed spectra of the XRT data accumulated in the first orbit
(150-2740 seconds since the burst trigger) and of the contributions from the two thermal
components derived from the BB+BB+PL model fitting.

{\bf Figure 2} The evolution of the power law photon indices ($\Gamma$) in the BB+PL
model (open circle) and the BB+BB+PL model (solid circle), respectively. While the BB+PL
case shows a strong evolution of $\Gamma$, no apparent evolution is revealed in the
BB+BB+PL case, with an average $\Gamma\sim 2.2\pm 0.4$.

{\bf Figure 3} The evolution of the blackbody emission radii ($left$) and temperatures
($right$) of the low-$T$ (open circles) and high-$T$ (solid circles) components,
respectively, derived from the time-resolved spectral analysis. The solid and dashed
lines are the regression lines for the high-$T$ and low-$T$ components, respectively. In
our spectral analysis we first divide the XRT data into 50 temporal segments. Each
segment covers $\sim 50$ seconds. We generally require that the spectral bins are $>100$
doe each segment. Some segments have a significantly smaller number of spectral bins. We
then merge these segments with their adjacent ones. We finally obtain 32 temporal
segments. The error bar in the time axis marks the time interval of each segment. We fit
the spectrum with the BB+BB+PL model. In the time-resolved spectral analysis the $N_{\rm
H}$ of the host galaxy is taken as $N_{\rm H}=0.64\times 10^{22}$ cm$^{-3}$ without
considering its temporal evolution.

{\bf Figure 4} The unabsorbed (a) and absorbed (b) light curves of the two thermal
components derived from the time-resolved spectral analysis with the BB+BB+PL model and
their comparisons with the light curve of the non-thermal component in 5-10 keV band (c).
The fluxes are in units of $10^{-9}$ erg cm$^{-2}$ s$^{-1}$.

\clearpage

\begin{figure*}
 \centerline{\hbox{\psfig{figure=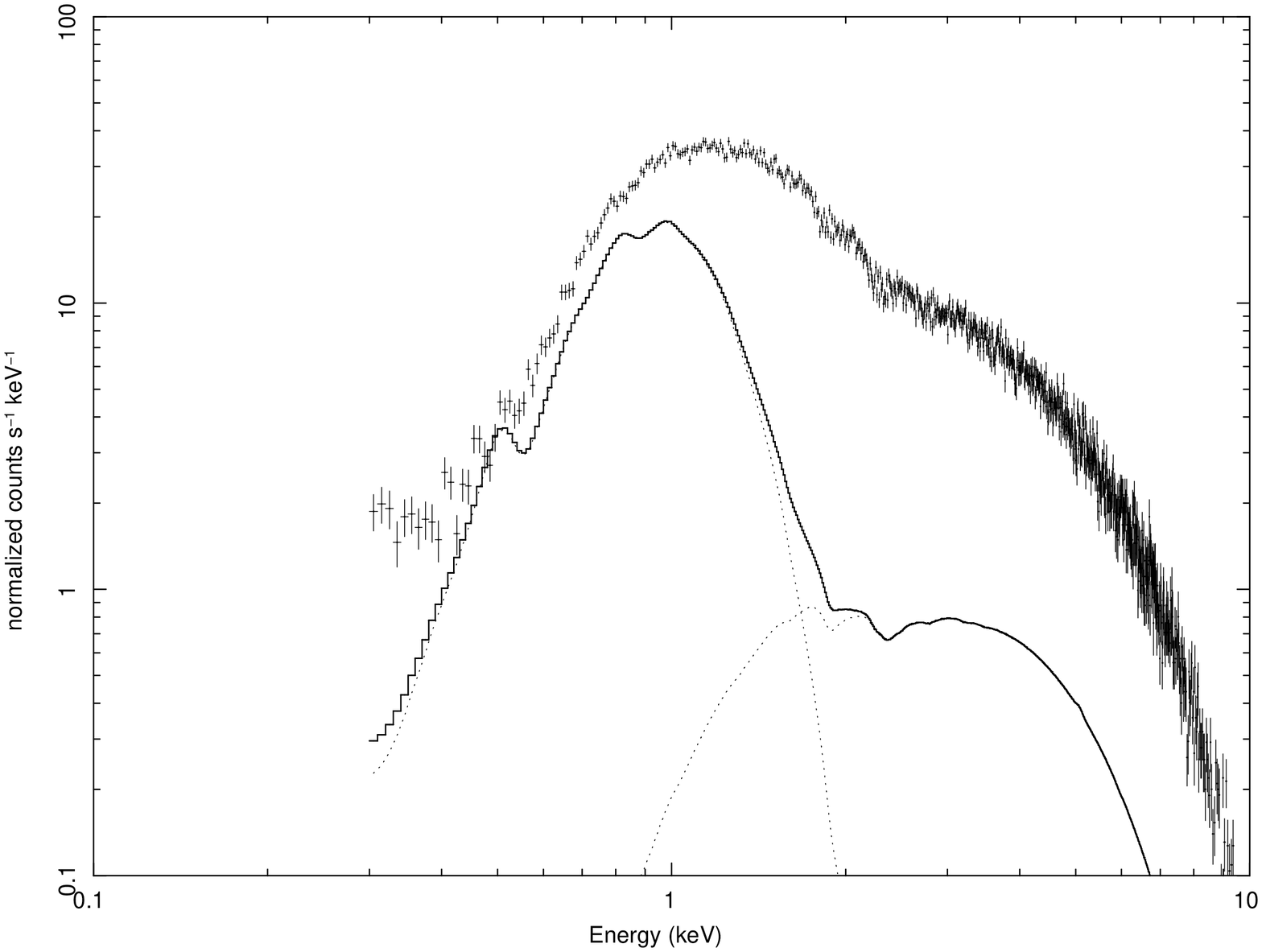,angle=-90,width=18cm}}}
\caption[]{}
%{\protect The observed spectra of the XRT data accumulated in the first orbit
%(150-2740 seconds since the burst trigger) and of the contributions from the two thermal
%components derived from the BB+BB+PL model fitting.}
\label{fig1}
\end{figure*}

\clearpage

\begin{figure*}
 \centerline{\hbox{\psfig{figure=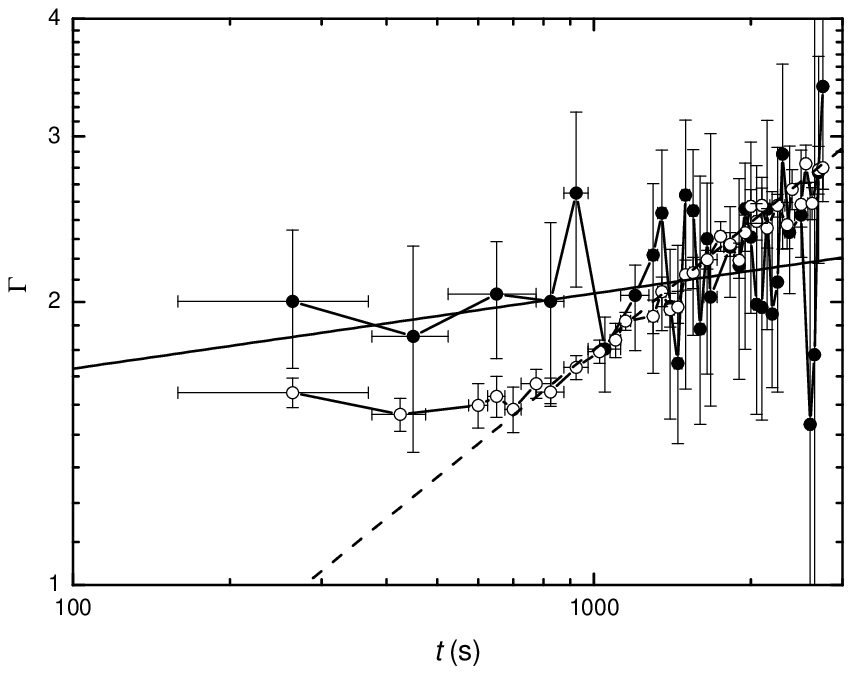,angle=0,width=18cm}}}
\caption[]{}
%\caption[]{\protect The evolution of the power law photon indices ($\Gamma$) in the BB+PL
%model (open circle) and the BB+BB+PL model (solid circle), respectively. While the BB+PL
%case shows a strong evolution of $\Gamma$, no apparent evolution is revealed in the
%BB+BB+PL case, with an average $\Gamma\sim 2.2\pm 0.4$.}
\label{fig2}
\end{figure*}

\clearpage
\begin{figure*}
 \centerline{\hbox{\psfig{figure=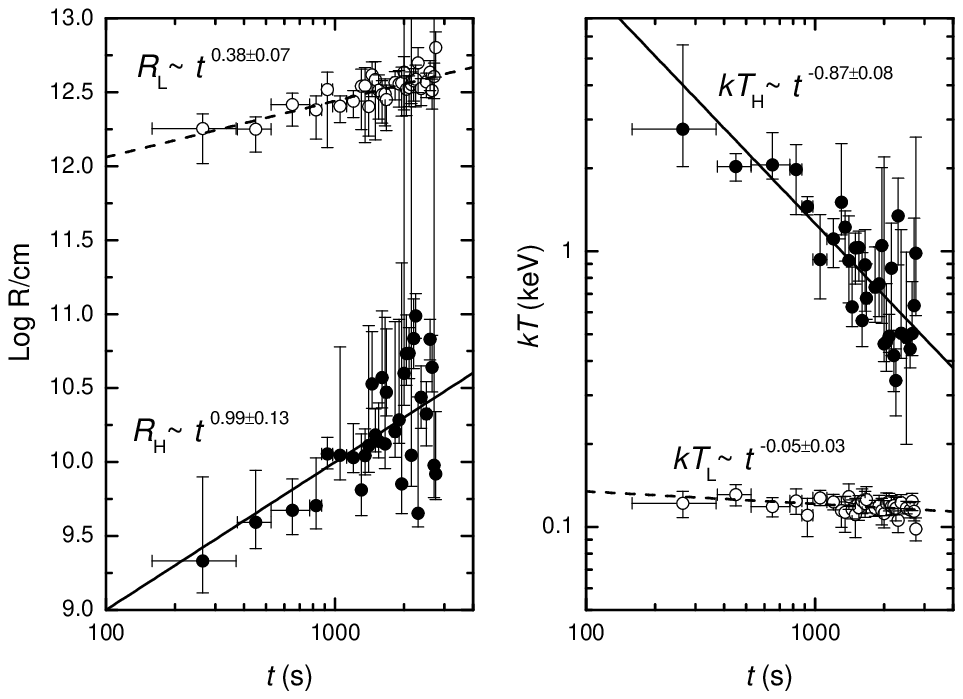,angle=0,width=18cm}}}
\caption[]{}
%\caption[]{\protect The evolution of the blackbody emission radii
%($left$) and temperatures ($right$) of the low-$T$ (open circles) and
%high-$T$ (solid circles) components, respectively, derived from the
%time-resolved spectral analysis. The solid and dashed lines are the
%regression lines for the high-$T$ and low-$T$ components,
%respectively. In our spectral analysis we first divide the XRT data
%into 50 temporal segments. Each segment covers $\sim 50$ seconds. We
%generally require that the spectral bins are $>100$ doe each segment.
%Some segments have a significantly smaller number of spectral bins. We
%then merge these segments with their adjacent ones.
%We finally obtain 32 temporal segments. The error bar in the time axis
%marks the time interval of each segment. We fit the spectrum with the
%BB+BB+PL model. In the time-resolved spectral analysis the $N_{\rm H}$
%of the host galaxy is taken as $N_{\rm H}=0.64\times 10^{22}$
%cm$^{-3}$ without considering its temporal evolution.}
\label{fig3}
\end{figure*}

\clearpage

\begin{figure*}
 \centerline{\hbox{\psfig{figure=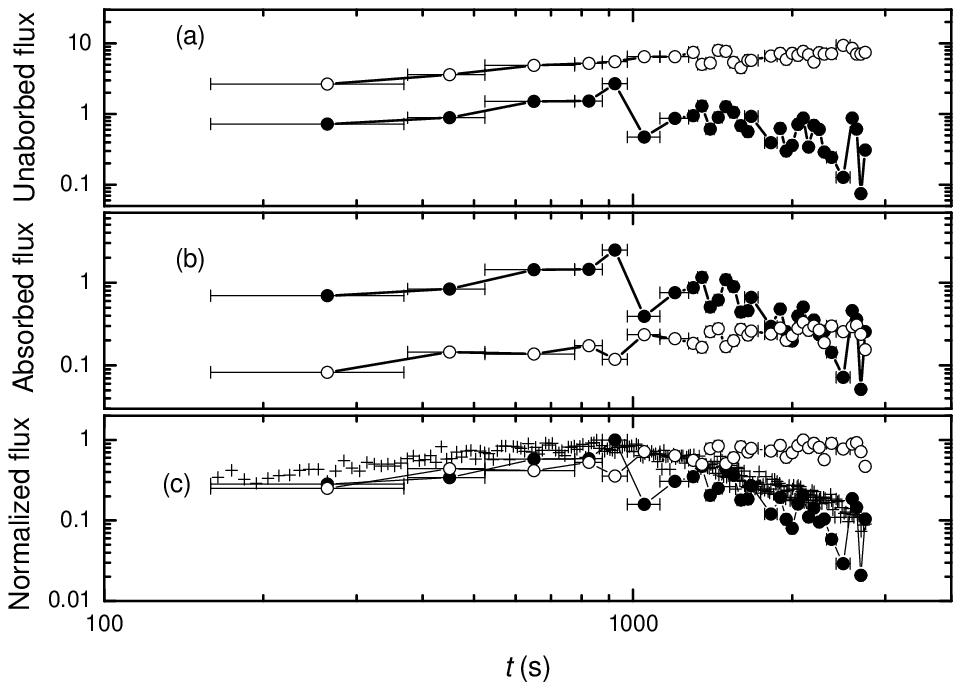,angle=0,width=18cm}}}
\caption[]{}
%\caption[]{\protect The unabsorbed (a) and absorbed (b) light curves of the two thermal
%components derived from the time-resolved spectral analysis with the BB+BB+PL model and
%their comparisons with the light curve of the non-thermal component in 5-10 keV band (c).
%The fluxes are in units of $10^{-9}$ erg cm$^{-2}$ s$^{-1}$.}
\label{fig4}
\end{figure*}

%\newpage
% \noindent{{\bf Supplementary Information} accompanies the paper on
%{\bf www.nature.com/nature}.}
%
%\section{Supplementary Table}
%
%\section{Supplementary Methods}
%
%\section{Supplementary Figure}

%\begin{acknowledge}
%This work was supported by NASA under grants NNG05GB67G,
%NNG05GH92G and NNG05GH91G (EWL, BZ \& ZGD), and the National Natural
%Science Foundation of China under grants 10463001
%(EWL) and 10221001 \& 10233010 (ZGD).
%\end{acknowledge}

\clearpage \vskip 0.5cm

\noindent{\bf Correspondence} and requests for materials should be
addressed to Enwei Liang (lew@physics.unlv.edu) and Bing Zhang
(bzhang@physics.unlv.edu).

\end{document}